\documentclass{pazhbeng}
\usepackage{graphicx}

\usepackage{lscape}
\usepackage{natbib}
\usepackage{threeparttable}
\usepackage[english]{babel}

\journalinfo{2013}{39}{6}{375}[388]
\submitted{21 December 2012}
\accepted{12 April 2013}

\def\ergs{erg s$^{-1}$}

\begin{document}

\title{\bf On Timing and Spectral Characteristics of the X-ray Pulsar 4U 0115+63:
Evolution of the Pulsation Period and the Cyclotron Line Energy
}

\author{
    P.~A.~Boldin\email{boldin.pavel@gmail.com}\address{1},
    S.~S.~Tsygankov\address{2,3,1},
    A.~A.~Lutovinov\address{1}\\
\bigskip
    {\it $^1$ Space Research Institute, Russian Academy of Sciences, Profsoyuznaya ul. 84/32, Moscow, 117997 Russia}\\
    {\it $^{2}$ Finnish Centre for Astronomy with ESO (FINCA), University of Turku, V\"ais\"al\"antie 20, FI-21500 Piikki\"o, Finland}\\
    {\it $^{3}$ Astronomy Division, Department of Physics, FI-90014 University of Oulu, Finland}\\
}

\begin{abstract}
An overview of the results of observations for the transient X-ray pulsar 4U 0115+63, a member
of a binary system with a Be star, since its discovery to the present day ($\sim$40 years) based
on data from more than dozen observatories and instruments is presented. A overall light curve and
the history of change in the spin frequency of the neutron star over the entire history of its observations,
which also includes the results of recent measurements made by the INTEGRAL observatory during the
2004, 2008, and 2011 outbursts, are provided. The source's energy spectra have also been constructed
from the INTEGRAL data obtained during the 2011 outburst for a dynamic range of its luminosities
($10^{37} - 7\times10^{37}$ \ergs). We show that apart from the fundamental harmonic of the cyclotron absorption
line at energy $\sim$11 keV, its four higher harmonics at energies $\simeq24, 35.6, 48.8$, and $60.7$~keV are
detected in the spectrum. We have performed a detailed analysis of the source's spectra in the
4-28 keV energy band based on all of the available RXTE archival data obtained during bright outbursts in
$1995-2011$. We have confirmed that modifying the source's continuum model can lead to the disappearance
of the observed anticorrelation between the energy of the fundamental harmonic of the cyclotron absorption
line and the source's luminosity. Thus, the question about the evolution of the cyclotron absorption line
energy with the luminosity of the X-ray pulsar 4U 0115+63 remains open and a physically justified radiation
model for X-ray pulsars is needed to answer it.
\end{abstract}
\section{INTRODUCTION}

The X-ray pulsar 4U 0115+63 has a long history of observations and is well suited for testing
models for the evolution of binary systems, models
for the structure and evolution of decretion disks
around Be stars, and models for the structure of emitting regions (accretion columns) in neutron stars.
The source was discovered in the UHURU satellite
survey \citep{giacconi1972,forman1978}.
Later on, this object was found in the Vela-5B satellite archival data since 1969 \citep{whitlock1989}.
Accurate measurements of its position in the sky
were subsequently made by the SAS-3, Ariel-V, and
HEAO-1 observatories \citep{cominsky1978,johnston1978}. These measurements allowed the
optical component in this binary system to be determined; it turned out to be a Be star called V635 Cas
\citep{kholopov1981} with an apparent magnitude
$V \approx 15.5$ and strong reddening \citep{johns1978,hutchings&crampton1981}.
A typical Be star is a hot ($T > 10000$~K), massive ($M > 5M_\odot$ ) O- or
B-type star with strong Balmer emission lines
detected in its spectrum. These lines originate in
the material outflowing from the star that forms the
so-called decretion equatorial disk \citep{slettebak1988,reig2011}.
This outflow of material is caused by a
high rotational velocity and a high light pressure on
the surface of the Be star. The distance to the system
4U 0115+63/V635 Cas is estimated to be $\approx 7$ kpc
\citep{negueruela&okazaki2001a}. The orbital parameters of the binary system were determined from
SAS-3 observations: the orbital period is $\approx24.3$ days,
the eccentricity is $e \simeq 0.34$, the projected semimajor
axis is $a_x \sin i \approx 140$~lt-s, and the mass function is
$f (M ) \simeq 5M_\odot$ \citep{rappaport1978}.

The pulse profile for the X-ray pulsar 4U 0115+63
has a simple double-peaked shape with a tendency
for the relative contribution from one of the peaks
to decrease with increasing energy, while at energies
above $\sim20$ keV the profile is essentially a single-peaked one.
The relative intensity of the peaks in
the profile also changes with the pulsar's intrinsic
luminosity. Such a behavior of the pulse profile is
typical for transients with Be-type companions \citep{lutovinov_tsygankov_2009}.
The properties of the
pulse profile for 4U 0115+63 have been studied in
detail in a number of works based on data from various
observatories \citep{johnston1978,leahy1991,lutovinov2000,tsygankov2007,ferrigno2011,sasaki2012}.

Two types of outburst activity are observed in X-ray transients that are members of binary systems
with Be stars. Type I outbursts have a strict periodicity modulated by periastron
passage of the neutron star and a moderate luminosity,
$10^{36} \lesssim L_x \lesssim 10^{37}$ \ergs.
Type II outbursts with luminosities
$L_x \gtrsim 10^{37} $ \ergs\ are difficult to predict due to their
less strict periodicity and much longer recurrence
periods. Survey missions periodically covering the
entire sky during a short time interval are required to
detect such outbursts, because Be transients spend
only a small fraction ($\lesssim 1\%$) of their history in type II
outbursts. The absence of a strict periodicity, the
long intervals between outbursts, and the source's
high luminosity are related to the outburst generation
process itself: the growth of a decretion disk around
the Be star, its subsequent ejection due to radiative
instability, and the capture of material by the neutron
star \citep{okazaki&negueruela2001}. The decretion
disk of the Be star becomes denser due to the disk
truncation by the tidal interaction of the neutron star
\citep{okazaki&negueruela2001}. This interaction
also gives rise to a gap between the periastron of
the neutron star orbit and the disk edge, which suppresses the type I outburst activity.

The transient X-ray pulsar 4U 0115+63 exhibits
mainly the type II outburst activity, giant outbursts
occurring with a periodicity of 3-5 years
\citep[see, e.g.,][]{wheaton1979,okazaki&negueruela2001}.
The outbursts are preceded by an optical brightening
of the system, suggesting the existence and growth
of a decretion disk around the Be star. The X-ray
luminosity of the object in quiescence is less than
$10^{34}$ \ergs\ \citep{campana2001}.

The source being studied is the first X-ray transient with a Be companion in the energy spectrum
of which an absorption line was detected at energy
$20$~keV \citep[\mbox{\em{HEAO-1 A4}}]{wheaton1979}
subsequently, it was interpreted as the first harmonic
of the fundamental cyclotron absorption line at energy $\approx11.5$~keV \citep{white83}.
At present, 4U 0115+63 is the only X-ray pulsar in the spectrum of which four
harmonics of the cyclotron absorption line, along with the fundamental one, have
been detected. In addition, an anticorrelation between
the position of the cyclotron absorption line and the
source's intrinsic luminosity was recorded with confidence for two bright transient pulsars (4U 0115+63
and V 0332+53). Such a dependence is observed both on long timescales
\citep[see, e.g.,][]{mihara2004,nakajima2006,nakajima2010,mowlavi2006,
tsygankov2006,tsygankov2007,tsygankov2010} and for variability on the scale of
the pulsation period \citep{klochkov2011b}. Such a behavior can be explained in
general terms within the classical model of accretion
column change with luminosity \citep{basko76}, which has been further developed recently
by \citep{poutanen2013}. Since data on the behavior of the cyclotron line
energy allow accretion column models to be tested, their accurate measurement is an
important task.

Here, we provide a complete historical overview
of the light curves and pulse period for 4U 0115+63
and present new results for the 2004, 2008, and
2011 outbursts obtained from INTEGRAL data.
In addition, we continue the study and interpretation of the behavior of the cyclotron absorption
line as the source's luminosity changes begun by
\cite{burnard1991,mihara2004,nakajima2006,tsygankov2007,li2012,muller2013}.
We discuss the luminosity -- fundamental cyclotron line harmonic
energy anticorrelation and show that the latter depends on the chosen model spectrum and,
for a certain shape of the continuum, the line energy is
practically constant.
In addition, based on the INTEGRAL
data obtained during the 2011 outburst, we have
constructed broadband spectra of the source for
various luminosity levels in which, apart from the
fundamental harmonic, four more higher harmonics
of the cyclotron absorption lines are detected.

\section{OBSERVATIONS AND DATA ANALYSIS}

To construct the light curve and the evolution of
the pulsation period for the X-ray pulsar 4U 0115+63,
we processed the observational data from more than
dozen observatories and instruments that have operated
or are operating in orbit from 1969 to the present days.
In particular: the Vela-5B satellite data were taken
from the HEASARC (High Energy Astrophysics
Science Archive Research Center) open archive; the
UHURU, Ariel-V, and GINGA satellite data were
digitized from the papers listed below (see the Section
“Light Curves and Evolution of the Pulsation Frequency”);
the GRANAT satellite data were retrieved
from the archive of the Department of High-Energy
Astrophysics at the Space Research Institute of the
Russian Academy of Sciences;
the BATSE data
(fluxes and pulsation frequencies) were digitized from
papers, except the February-March 1999 outburst
whose data are available in the BATSE%
\footnote{http://www.batse.msfc.nasa.gov/batse/pulsar/data/}
archives;
the RXTE/ASM data were taken from the official site
of the ASM/RXTE\footnote{http://xte.mit.edu/asmlc/ASM.html}
development team; the team
of the Swift/BAT all-sky monitor provides archival
data and a list of transients via the corresponding
website\footnote{http://swift.gsfc.nasa.gov/docs/swift/results/transients/};
the MAXI monitor data were taken from
the official site\footnote{http://maxi.riken.jp/top/index.php?cid=1\&disp\_mode=
source}.
The INTEGRAL observatory \citep{winkler03}
performed series of observations for the X-ray pulsar
4U 0115+63 during the 2004 \citep{tsygankov2007},
2008 \citep{li2012}, and 2011 outbursts. The results
of our timing analysis of these data were also used
to construct the light curve and the evolution of the
pulsation period. The OSA-10%
\footnote{http://isdc.unige.ch} standard software was
used for processing the JEM-X data and for our
timing analysis of the data from the ISGRI detector
of the IBIS telescope. The software developed at the
Space Research Institute of the Russian Academy of
Sciences \citep{krivonos10} was used to construct
the light curves and spectra from ISGRI/IBIS data.
Here, we also analyzed the data from all RXTE
observations of 4U 0115+63 \citep{bradt1993}. In
particular, we used the PCA data obtained during the
1999, 2000, 2004, 2008, and 2011 outbursts for our
spectral analysis. The standard set of programs from
the FTOOLS version 6.12 package was used for the
RXTE data processing and the final analysis of all the
remaining observations.

\section{RESULTS OF TIMING ANALYSIS}
\subsection{Light Curves and Evolution of the Pulsation Frequency}

\begin{table*}
  \centering
  \begin{threeparttable}[b]
    \caption{Known significant outbursts of the X-ray transient pulsar 4U 0115+63 since its discovery to the present day as
revealed by the listed observatories and instruments. The first column gives the corresponding parts of Fig. 2, where the
outburst light curves and pulsation frequencies are presented. For details, see the text
    \label{table:flares}}
    \begin{tabular}{ l p{4.0cm} l l l p{3.5cm} l }
    \hline \hline
& Date & Begin & End & $F_\mathrm{3-100\,\text{keV}}$, & Instruments & References\tnote{1}\\
& & MJD & MJD & Crab && \\[1mm]
\hline
(a) & 1969-1971\tnote{2} & $\sim$40000 & $\sim$41000 & 1.0 & {\em Vela 5B, UHURU} & 1, 2\\
    & August 1974 & 42265 & 42295 & $>$1.5 & {\em Vela 5B} & 2 \\
(b) & December 1977 & 43500 & 43550 & $>$1.0 &\mbox{{\em Vela-5B}, {\em SAS-3}}, {\em Ariel-V}, {\em HEAO-1} & 2, 3, 4, 5 \\
    & December 1980 & 44589 & & $>$0.2 & {\em Ariel-VI} & 6 \\
    & February-March 1987 & 46835 & 46865 & 0.2 & {\em GINGA} & 7 \\
(c) & February 1990 & 47935 & 47950 & 0.2 & {\em GINGA, GRANAT} & 7 \\
(d) & April-May 1991 & 48360 & 48375 & 0.07\tnote{3}  & {\em BATSE} & 8, 9 \\
(e) & May-June 1994 & 49480 & 49530 & 0.13\tnote{3}  &  {\em BATSE} & 10 \\
(f) & November-December 1995 & 50040 & 50090 & 0.7 (0.14\tnote{3} ) & {\em BATSE, GRANAT} & 10, 11, 12 \\
(g) & August-November 1996\tnote{2} & $\sim$50250 & $\sim$50500 & 0.1 & {\em RXTE/ASM} & \\
(h) & February-March 1999 & 51230 & 51280 & 0.6 (0.2\tnote{3} ) & {\em RXTE/ASM}\tnote{ 4}, {\em BATSE} & 13 \\
(i) & September-October 2000 & 51780 & 51830 & 0.4 & {\em RXTE/ASM}\tnote{ 4} & \\
(j) & September-October 2004 & 53240 & 53300 & 0.6 & {\em RXTE/ASM}\tnote{ 4}, {\em \mbox{INTEGRAL}}\tnote{ 5} &   \\
(k) & March-April 2008 & 54530 & 54580 & 0.5 & {\em RXTE/ASM}\tnote{ 4}, {\em Swift/BAT}\tnote{ 6}, {\em \mbox{INTEGRAL}}\tnote{ 5} &  \\
(l) & May-June 2011 & 55720 & 55755 & $\sim 0.8$ & {\em RXTE/ASM}\tnote{ 4}, MAXI, {\em Swift/BAT}\tnote{ 6}, {\em \mbox{INTEGRAL}}\tnote{
5} &  \\
\hline
    \end{tabular}
    \begin{tablenotes}
    \item [1]
    (1) \cite{giacconi1972}
    (2) \cite{whitlock1989}
    (3) \cite{cominsky1978}
    (4) \cite{rose1979}
    (5) \cite{johns1978}
    (6) \cite{ricketts1981}
    (7) \cite{tsunemi1988}
    (8) \cite{cominsky1994}
    (9) \cite{bildsten1997}
    (10) \cite{wilson1994}
    (11) \cite{finger1995}
    (12) \cite{sazonov1995}
    (13) \cite{wilson1999}
    \item [2] Series of type I outbursts.
    \item [3] Pulsed flux.
    \item [4] Data were provided by the {\em RXTE/ASM}.
    \item [5] This paper.
    \item [6] Data were provided by the {\em Swift/BAT}.

    \end{tablenotes}
  \end{threeparttable}
\end{table*}
\begin{table*}
  \centering
    \caption{ History of orbital period measurements for 4U\,0115+63 \label{table:porb} }
    \begin{tabular}{ l l l l }
        \hline \hline
    T, MJD & $P_\mathrm{orb}$,~days & Observatory & Reference \\
        \hline
    $40963.81\pm0.40$ & $24.3149\pm0.0044$ & UHURU & \cite{kelley1981} \\
    $43540.951\pm0.006$ & $24.309\pm0.021$ & SAS-3 & \cite{rappaport1978} \\
    $44589$ & $24.3155\pm0.0002$ & Ariel 6 & \cite{ricketts1981} \\
    $47942.224\pm0.004$ & $24.31643\pm0.00007$ & GRANAT & \cite{lutovinov2000} \\
    $49279.2677\pm0.0034$ & $24.317037\pm0.000062$ & BATSE & \cite{bildsten1997} \\
    $53243.038\pm0.051$ & $24.3174\pm0.0004$ & RXTE & \cite{raichur2010} \\
    \hline

    \end{tabular}
\end{table*}

Our investigation was naturally begun with a
study of the light curves for 4U 0115+63 based on the
observational data from X-ray monitors and observatories over the last 40 years, from the Vela-5B archival
data to the MAXI and Swift/BAT data
(see Table 1). Figure 1 presents a complete historical
light curve of the X-ray pulsar 4U 0115+63; Fig. 2
(upper panels) presents the same light curves but
with a better time resolution. Since the observations
were carried out with different instruments in different
energy bands, to present a uniform light curve, the
results of these observations were recalculated to the
3-100 keV energy band as follows: the flux measured
from the pulsar 4U 0115+63 in the instrument's band
is known from the observations; given
the shape of the pulsar's spectrum and assuming it to be constant,
we can find the flux from the source in the energy
band of interest normalized to the Crab flux in the
same energy band. To recalculate the fluxes, we took
the spectral shape for 4U 0115+63 in the form of a
power law with a high-energy cutoff typical of X-ray
pulsars \citep{white83}:
\begin{equation}
f(E)=AE^{-\Gamma}\times \left\{ \begin{array}{ll}
1 & \mbox{($E \leq E_{cut}$)}\\
exp^{-(E-E_{cut})/E_{fold}} & \mbox{($E > E_{cut}$),} \end{array} \right.
\end{equation}
with parameters $E_\mathrm{cut} = E_\mathrm{fold} = 8$ keV and $\Gamma = 0.1$.
Thus, all of the fluxes given below refer to the 3-100 keV energy band, unless stated otherwise.

To make Figs. 1 and 2 more demonstrable, we cleaned the light
curves of the measurements where the flux did not
exceed its measurement error and omitted the flux
measurement errors themselves. The lower panels in
Fig. 2 present the evolution of the neutron star spin
frequency in which its increase (spin-up of the neutron star) over the entire
history of the source's observations is clearly traceable.
A correlation between
the outbursts and the increase in neutron star spin
frequency can also be seen. This spin-up of the neutron
star is related to the positive angular momentum
transferred by the infalling material from the accretion
disk \citep[see, e.g.,][]{ghosh&lamb1979}. It can also be
clearly seen that the neutron star spins down between
the outbursts. This may suggest that the neutron
star magnetosphere interacts with the surrounding
material from the stellar wind of the Be star, for example, through
the propeller mechanism \citep{illarionov&sunyaev1975,shakura75}. For a clear
demonstration of the described effects, the dashed lines
in the lower panels of Fig. 2 indicate the last known
neutron star spin frequency before the onset of an
outburst. It is also interesting to note
the neutron star continues to slow down for some
time at the onset of many outbursts and only then
does it begin to speed up.

As has been said above, the X-ray pulsar 4U 0115+63
was discovered by the UHURU (2-6 keV) satellite
during its type I outburst with a peak flux of 0.2 Crab
(2-6 keV) and was initially named 2U 0115+63.
Subsequently, this object was found in the Vela-5B
archival data, where several type I outbursts were
detected (August 1969, January 1970, August 1970,
January 1971; Fig. 2a, \cite{whitlock1989}). Thereafter,
the type I activity essentially ceased and,
starting from the outburst in January 1971 \citep{whitlock1989},
4U 0115+63 has exhibited mainly the
type II activity, giant outbursts with a recurrence
period of about three years (Fig. 1). The outburst
in August 1974 was detected only by the Vela-5B
satellite. We took the spin frequencies calculated
from UHURU data from \cite{ricketts1981}and
\cite{kelley1981} and those calculated from Vela-5B data from \cite{whitlock1989}(indicated by the
diamonds in the lower panel of Fig. 2a).

The outburst in December 1977-January 1978
was detected by the instruments of the SAS-3
\citep{cominsky1978}, Ariel-V \citep{rose1979},
and HEAO-1 \citep{johns1978} observatories. \cite{rose1979} investigated
phase-resolved spectral
characteristics and showed them to change greatly
with pulse phase. The pulsation frequency determined by \cite{ricketts1981}
from SAS-3 observations is indicated by the diamond in the lower panel
of Fig. 2b. The outburst with an intensity of 0.2 Crab
(3-6 keV) in December 1980 was detected only by
the Ariel-VI satellite \citep{ricketts1981}. The
orbital parameters calculated from these observations
allowed the mass of the Be star to be constrained,
$M_c \lesssim 25M_\odot$. No outbursts were detected from 1980
to 1987 probably because there were no suitable X-ray monitors in orbit at this time.

The next small (with an intensity up to 0.18 Crab)
outburst was detected by the ASM instrument (1-20 keV energy band) of the GINGA telescope in
February-March 1987 \citep{tsunemi1988}.
These observations confirmed the $\sim3$-year periodicity
of outbursts. The outburst in February 1990 (Fig. 2c)
with a peak flux of about 0.4 Crab (1-20 keV)
was also observed by the GINGA satellite \citep{makino1990}.
The apsidal motion in the system
$\dot \omega = 0\overset{\circ}{.}030 \pm 0\overset{\circ}{.}016$~yr$^{-1}$.
was determined from these data \citep{tamura1992}. In addition, observations
with the LAC instrument of the GINGA observatory
allowed one to study the source's spectra in the 1-37 keV energy band and to detect two absorption
lines at energies $\approx 12.5$ and $\approx22.6$ keV.
This confirmed the previous conclusion about the
simultaneous presence of two cyclotron absorption
line harmonics in the pulsar spectrum. Concurrently
with the GINGA satellite, a series of observations
was performed for the source with the ART-P (3-30 keV) telescope of the GRANAT observatory on
February 18, 19, and 22, 1990 \citep{gilfanov1991,lutovinov1994}.
These observations allowed one to refine the epoch of periastron passage,
$\tau(\mathrm{MJD}) = 47942.224 \pm 0.004$
and to measure the orbital period of the binary system,
$P_\mathrm{orb} = 24.31643 \pm 0.00007$~days. Two cyclotron absorption lines at
energies $\approx12.10$ and $\approx22.24$ keV were also detected
in the spectra of the pulsar 4U 0115+63 \citep{lutovinov2000}.
The measured period of the pulsar was $P = 3.61461 \pm 0.00001$
(indicated by the diamond in the lower panel of Fig. 2c).

The small outburst in April-May 1991 was detected by the BATSE instrument
\citep{cominsky1994,bildsten1997} onboard the Compton-GRO observatory. The peak pulsed flux during the
outburst was 0.07 Crab (20-50 keV) and its duration
was about 10 days (see Fig. 2d).

The outburst in May-June 1994 with a duration
of 50 days and a peak pulsed flux of 0.13 Crab and
the outburst in November-December 1995 with a
duration of 30 days and a peak pulsed flux of 0.14 Crab
were also detected only by the BATSE instrument
\citep[see Figs. 2e and 2f, respectively]{wilson1994}.
The spin period of the neutron star and its evolution
presented in the lower panels of these figures were
determined from these data \citep{bildsten1997}. The
outburst in November 1995 was also detected by the
GRANAT/WATCH instrument with a peak flux of
0.7 Crab in the 8-20 keV energy band \citep{sazonov1995}.

Shortly after the launch of the RXTE observatory
in December 1995, the All-Sky Monitor (ASM) (the
operating energy range 1.3-12 keV) detected a series
of weak type I outbursts from August to November
1996 (Fig. 2g). The peak flux during these outbursts
was 0.1 Crab in the instrument's band. The pulsation
frequency determined from Compton GRO data was
276.667 mHz \citep[indicated by the
diamond in the lower panel of Fig. 2g]{scott1996}.

The outburst in February-March 1999 with a
peak flux of 0.8 Crab (1.3--12 keV) was detected by
both RXTE/ASM and BATSE with a pulsed flux of
0.2 Crab in the 20-50 keV energy band. The outburst
lasted for about 50 days (see Fig. 2h). The next
outburst in September-October 2000 was already
observed solely by RXTE/ASM; the peak flux was
about 0.38 Crab (1.3-12 keV) and the duration was
40 days (Fig. 2i).

In September 2004, the onset of another outburst was
recorded by the INTEGRAL observatory
\citep{lutovinov2004}. Subsequently, the outburst
was observed with the instruments of the RXTE and
INTEGRAL observatories
with a peak flux of 0.5 Crab in the 20–60 keV energy band.
Variations of the neutron star spin frequency during the outburst
were determined from INTEGRAL data. The results
of these measurements are presented in the lower
panel of Fig. 2j, along with those from RXTE data
\citep[indicated by the diamond]{raichur2010}.
With the launch of the Swift observatory in 2004,
an all-sky monitoring in hard X-rays became possible
and the succeeding 4U\,0115+63 outbursts' were detected already
in a wide energy range. In particular, the peak fluxes
recorded during the outburst in March-April 2008
were 0.4 Crab for RXTE/ASM and 0.6 Crab for
Swift/BAT (15-50 keV). At the same time, INTEGRAL observations
of the pulsar 4U 0115+63
were performed. They allowed the spin period of the
neutron star to be determined (see Fig. 2k, the lower panel).

The last known outburst from 4U 0115+63 occurred in May-June 2011
and was observed by the
Swift/BAT and INTEGRAL observatories as well as
by the MAXI (2-20 keV) monitor. The peak flux
during the outburst was $\simeq0.4$ Crab from Swift/BAT
data and $\simeq1.0$ Crab from MAXI data in the corresponding
bands of the instruments. The observations
performed with the INTEGRAL telescopes at our
request allowed the spin frequency of the neutron star
to be measured; the results of these measurements are
shown in the lower panel of Fig. 2l. It also shows the
evolution of the neutron star spin frequency during the
outburst determined using data from the Gamma-ray
Burst Monitor (GBM) of the Fermi%
\footnote{Fermi/GBM, http://f64.nsstc.nasa.gov/gbm/} observatory.

    \begin{figure*}[p]
        \centering
        \includegraphics[width=0.5\textwidth,angle=0]{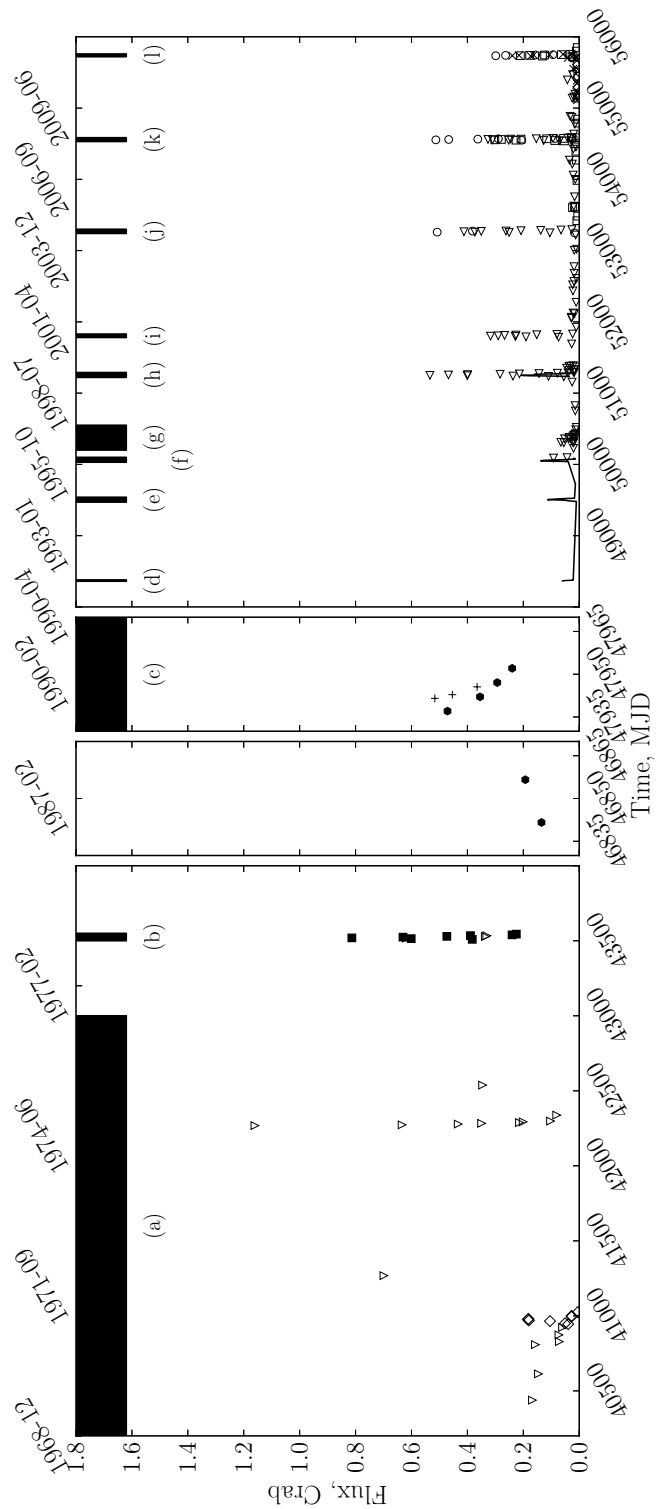}
        \caption{
Overall historical light curve of the X-ray pulsar 4U\,0115+63. The source
intensity in the 3-100 keV energy band is plotted on the Y-axis
in units of the Crab intensity (see text for details).
Most significant outbursts are emphasized by the grey color and, together with
the evolution of the neutron star spin frequency, are shown in details
on top and bottom panels of Fig.  \ref{pic:history}, respectively.
Different symbols mark results of observations by different observatories:
open triangles faced down --- {\em Vela-5B}, open diamonds --- {\em UHURU},
filled squares --- {\em Ariel-V}, filled hexagons --- {\em GINGA},
crosses --- {\em GRANAT}, solid line --- {\em BATSE} of {\em Compton GRO} (see
text), open triangles faced left --- {\em RXTE/ASM}, open circles ---
{\em INTEGRAL}, open squares --- {\em Swift/BAT}, 'x' symbols --- {\em MAXI}.
\label{pic:lc}
}
    \end{figure*}

\pagebreak
\newpage

    \begin{figure*}
      \centering
      \hspace{1cm}\hbox{
      \includegraphics[width=0.35\textwidth,angle=0]{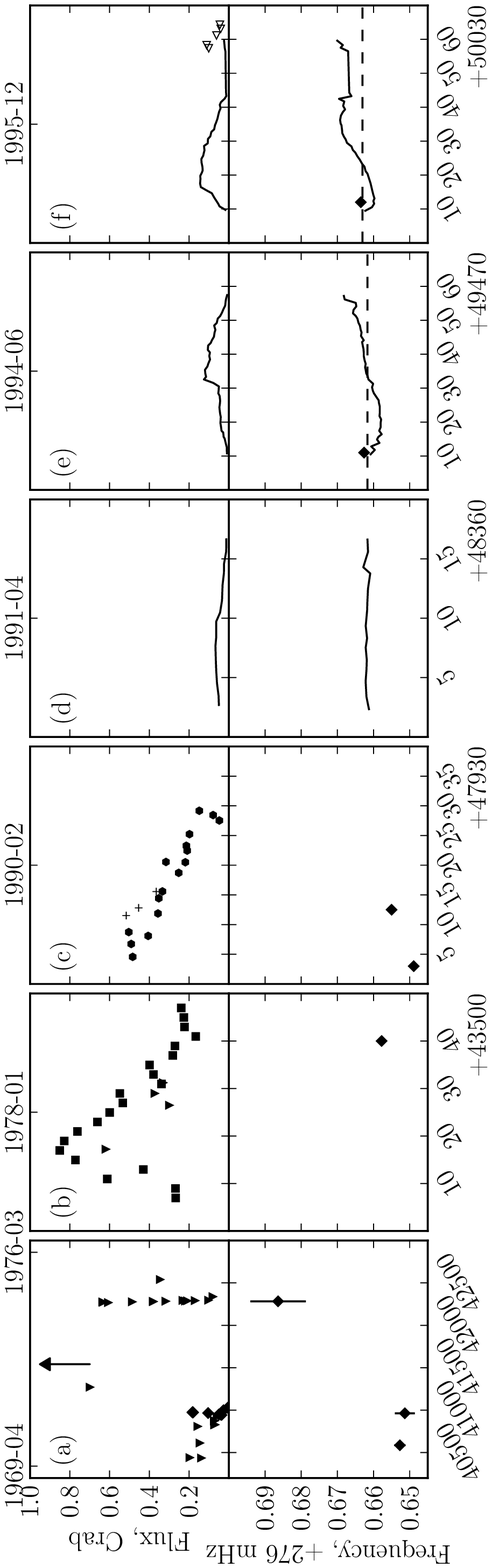}
      \hskip1em
      \includegraphics[width=0.35\textwidth,angle=0]{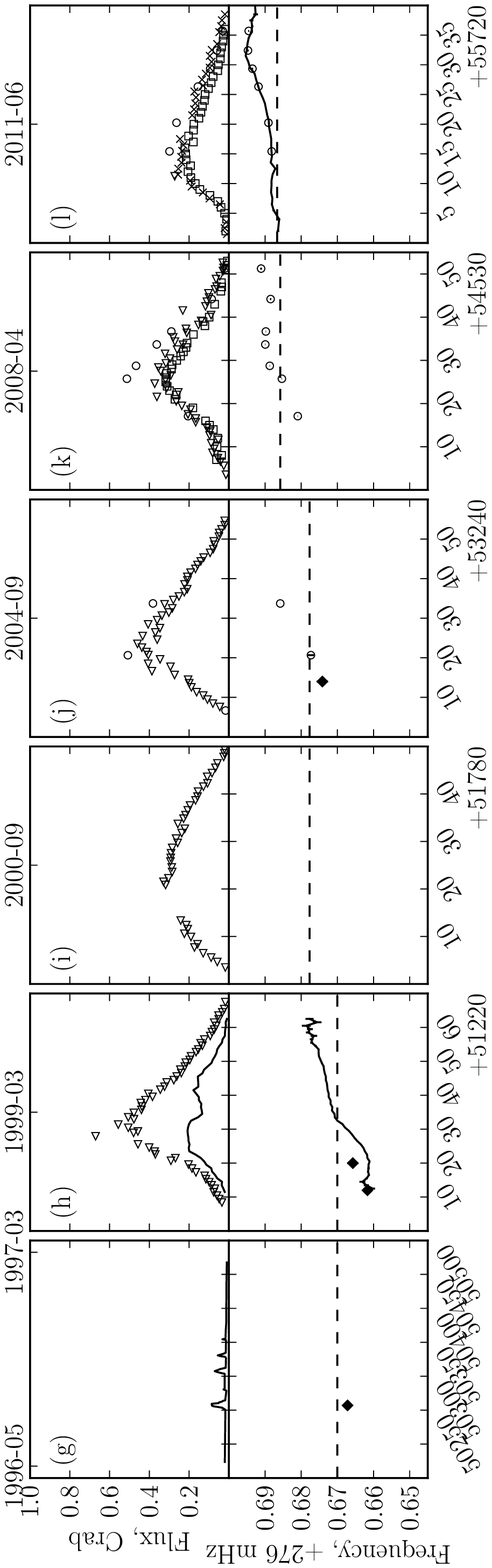}
      }
      \caption{
Variations of the flux (upper panels) and pulse frequency (lower panels)
of the X-ray pulsar 4U\,0115+63 during the most significant outbursts.
Symbols on upper panels are the same as for Fig.~\ref{pic:lc},
except for panel {\it g}, where results from the {\em RXTE} observatory are
shown by solid line.
Symbols on lower panels are the same as on upper ones with following
additions:
diamonds denote results of measurements taken from telegrams and other
information sources
(see text for details); solid line on panel {\it l} represents results of the
{\em Fermi/GBM} measurements.
Dashed horizontal lines on lower panels show the last known neutron stars'
spin frequency before
the start of the new outburst.
\label{pic:history}
       }
    \end{figure*}

\newpage
\pagebreak

\subsection{Determination of the Pulse Period from INTEGRAL data}

The spin frequencies of the neutron star
in the system 4U\,0115+63/V635 Cas
determined here from INTEGRAL data are shown in the
lower panels of Figs. 2j-2l. The spin frequencies
and their errors were calculated as follows. After
the correction to the barycenters of the Solar system
and the binary system, each light curve was analyzed
by the epoch folding technique \citep{leahy1983}.
The derived $P$--$\chi^2$ relationship was then fitted by
a Gaussian.
In this way, the pulsation period was
determined for the original light curve.
The errors in the period were estimated for each
light curve by the so-called bootstrap method (for a
brief description, see \cite{lutovinov2012}. First, we
generated the light curves in which the count rate at
each instant of time was determined from the formula
$r'_i = r_i + \gamma \sigma_{r_i}$, where $r'_i$
i is the count rate at the $i$th
point of the new light curve, $r_i$ is the same for the original light curve,
$\gamma$ is a quantity uniformly distributed
in the interval (--1; 1), and $\sigma_{r_i}$ is the measurement
error of the flux for the $i$th point. Subsequently, as
in the case of the original light curve, we analyzed
the derived light curve by the epoch folding technique
and determined the most probable pulsation period for
it. The result of repeating such an analysis $N \simeq 1000$
times is a vector of periods $P_i$ for different realizations
of the original light curve. The mean of the sample
$\langle P \rangle =\sum\limits^N_i P_i / N$
in this case is the most probable period and
standard deviation
$\sigma_\mathrm{P} = \sqrt{ \sum\limits^N_i \left(P_i - \langle P \rangle\right)^2 / N }$
is the estimate of the error of the period.

\subsection{Optical Observations}

The optical observations of 4U 0115+63/V635 Cas
carried out by \cite{kriss1983} allowed its X-ray
outburst to be predicted for the first time based on
the brightening of the optical star. This additionally
confirmed that the object V635 Cas was correctly
identified as a normal companion in the system. The
model of direct accretion from a stellar wind was
rejected based on the measured delay between the
optical brightening and the X-ray outburst.

The optical observations of the companion star
V635 Cas performed from 1985 to 1990 at the Wise
Observatory corroborated the previously reached
conclusion that an X-ray outburst is preceded by
an optical brightening \citep{mendelson1987,mendelson1991}.
In particular, the 1987 X-ray outburst was
preceded by the system's optical brightening from
October 1986 to February 1987. The 1990 outburst
was preceded by a large optical brightening that
had continued until the completion of the program
of observations in March 1990.
It should be noted
that the object's optical brightening in 1988 was not
accompanied by an X-ray outburst \citep{mendelson1991}.

Optical and infrared observations
of the source 4U 0115+63 were carried
out from 1999 to 2006 \citep{reig2007}. Their goal
was to test the models of a decretion disk around
a Be star proposed previously by \cite{negueruela&okazaki2001a}.
A change in the $H_\alpha$ line profile was
recorded, suggesting that it is the disk formation and
loss that affect the periodicity of type II outbursts in
4U 0115+63 \citep{negueruela&okazaki2001a}.

\subsection{Orbital Parameters}

As has already been said in the Introduction, the
orbital parameters of the pulsar 4U 0115+63 were
first calculated from SAS-3 observations \citep{rappaport1978}.
In particular, the orbital period of
the system calculated from these data was $24.309 \pm 0.021$~days.
Subsequent observations allowed this
estimate to be improved. The orbital parameters
calculated from BATSE data are used in most papers
for correction to the barycenter of the binary system
4U 0115+63 \citep{bildsten1997}. We used these
data here to calculate the pulsation period from INTEGRAL data;
the results are presented in the lower
panels of Figs 2j-2l and are described in the previous
section. On the whole, however, long-term observations of
4U 0115+63 and regular measurements of
its orbital parameters suggest that the orbital period
in the binary system increases (see Table 2). This is
diametrically opposite to the observed changes of the
orbital period in ordinary (not Ве) high-mass X-ray
binaries, where it decreases (Falanga et al. 2013, in
preparation).

The source's high brightness during outbursts
and its relatively short pulsation period allowed the
change of not only the orbital period in the binary
system but also other parameters to be traced.
In particular, using UHURU data and results from
\cite{rappaport1978,kelley1981} obtained
a constraint on the motion of the periastron for the
neutron star orbit,
$\dot \omega \lesssim
2\overset{\circ}{.}1$~yr$^{-1}$.
The motion of the
periastron for the neutron star orbit determined later
from GINGA data during the February 1990 outburst
$\dot \omega = 0\overset{\circ}{.}030 \pm 0\overset{\circ}{.}016$~yr$^{-1}$
\citep{tamura1992}.
More recent measurements based on RXTE data give
$\dot \omega=0\overset{\circ}{.}06 \pm 0\overset{\circ}{.}02$~yr$^{-1}$
\citep{raichur2010}.

\section{SPECTRUM}
\subsection{Dependence of the Fundamental Cyclotron Line
            Energy on Continuum Model}

There exist two main approaches to describing the
spectra of the radiation originating in the accretion
columns of X-ray pulsars. The first approach consists
in a physically proper modeling of the spectrum formation
processes and leads to physically
justified but often excessively complex models.
However, even using a more or less physically
proper model does not allow the source's observed
spectrum to be completely described. In particular,
\cite{ferrigno2009} attempted to use the model by
\cite{beckerwolff2007} in analyzing the spectra of
the outburst observed from 4U 0115+63 in March
1999 by the BeppoSAX observatory. However, to
completely describe the observed spectra, they had to
modify the model and to add a Gaussian
component at energy $\sim$9 keV to it. The latter is required to remove
the observed difference between the
observational data and the model and, as the authors
themselves admitted, carries no any physical meaning.
It should be noted that a similar artificial component
was also required for the authors to describe the
spectra by a phenomenological model. An interesting
consequence of adding a Gaussian component to
describe the source's spectrum is the absence of any
change in the energy of the fundamental cyclotron
line harmonic with the source's intensity.

The second approach consists in a phenomenological description of the spectrum,
i.e., in choosing
a sufficiently good model whose rigorous physical
justification is ignored. One of two similar models are
mainly used for a phenomenological description of the
spectra of X-ray pulsars: (1) a power-law spectrum
with a cutoff $\sim E^{-\Gamma} \exp\left(-\frac{E}{E_\mathrm{cut}}\right)$
and (2) a powerlaw spectrum with a high-energy cutoff (see eq. 1 above).

As has already been said above, several harmonics
of the cyclotron absorption line are detected in the
spectrum of the X-ray pulsar 4U 0115+63, with the
energy of the fundamental harmonic roughly corresponding to
$E_\mathrm{cyc,0} \approx 12B_{12}$~keV, where $B_{12}$ is the
magnetic field strength of the neutron star in units
of $10^{12}$~G. The width of these lines is determined by
many factors: the spectrum averaging over the pulse
profile, the high temperature of the scattering region,
the natural quantum broadening, the dispersion of
the magnetic field strength over the scattering region,
etc.

There exist several models that describe the shape
of the cyclotron absorption lines. Here, we used the
multiplicative {\em cyclabs} model from the {\sc XSPEC} package
\begin{equation}
M(E) = \exp\left[-\tau
            \frac{\left(W_\mathrm{f}E/E_\mathrm{cycl}\right)^2}%
            {\left(E - E_\mathrm{cycl}\right)^2 + W_\mathrm{f}^2}
        \right]
\end{equation},
where $\tau$ is the
depth of the cyclotron absorption line, $E_\mathrm{cycl}$ (in keV)
is the energy of the cyclotron line center, and $W_f$ (in
keV) is the cyclotron line width.
In general, the positions of the cyclotron absorption features in the pulsar
spectrum depend on the apparent geometry, i.e., on
the pulse phase, because the scattering cross section
depends on the angle between the photon direction
and the direction of the magnetic field vector.
However, here we disregarded this effect for a simple analysis and
studied the pulsar spectra averaged over the pulse profile.

The X-ray pulsar 4U 0115+63 has the longest
history of searching for and detecting cyclotron absorption lines
(for the discovery and interpretation
of the fundamental and first harmonics, see above).
Based on RXTE data, \cite{heindl1999} detected
a second harmonic at energy $\approx33.56$ keV at the
descending part of the smaller peak in the pulse profile,
which provides evidence for different formation
regions of these lines. A third harmonic in this source's spectrum
was detected by \cite{santangelo1999} at energy 49.5 keV based
on BeppoSAX data and by \cite{tsygankov2007} at energy 44.93 keV based on INTEGRAL data. The
detection of a fourth harmonic in the source's spectrum was reported by \cite{ferrigno2009} based on
BeppoSAX data and by \cite{muller2013} based
on RXTE data. However, the harmonic energies
reported by the authors differ significantly, $\simeq53$ and
$\simeq60$ keV, respectively.

A change of the cyclotron line energy with luminosity has been
found for the first time for the pulsar
4U 0115+63 among all X-ray Be transients.
In particular, \cite{mihara2004} did not detect the two
expected cyclotron lines at $\sim$11 and $\sim$22 keV in the
descending part of the weak
($L_x = 2\times10^{37}$ \ergs)
outburst in March 1991 in the LAT/GINGA spectra.
Instead of this pair, they detected one strong absorption line
at energy $\sim$16 keV. Further studies based on
RXTE data \citep{nakajima2006} showed the energy
of the fundamental cyclotron line harmonic to change
from $\sim$11 to $\sim$16 keV at the source's luminosity
$L_x < 4-5\times10^{37}$ \ergs.
Such a behavior was subsequently confirmed by \cite{tsygankov2007}. Apart
from the variability of the cyclotron absorption line
energy on long time scales, the pulsar 4U 0115+63
also exhibits variability of the cyclotron line energy
from pulse to pulse \citep{klochkov2011b}, which
confirms the anticorrelation between its energy and
luminosity.

\begin{figure*}
    \centering
    \includegraphics[width=1.0\textwidth]{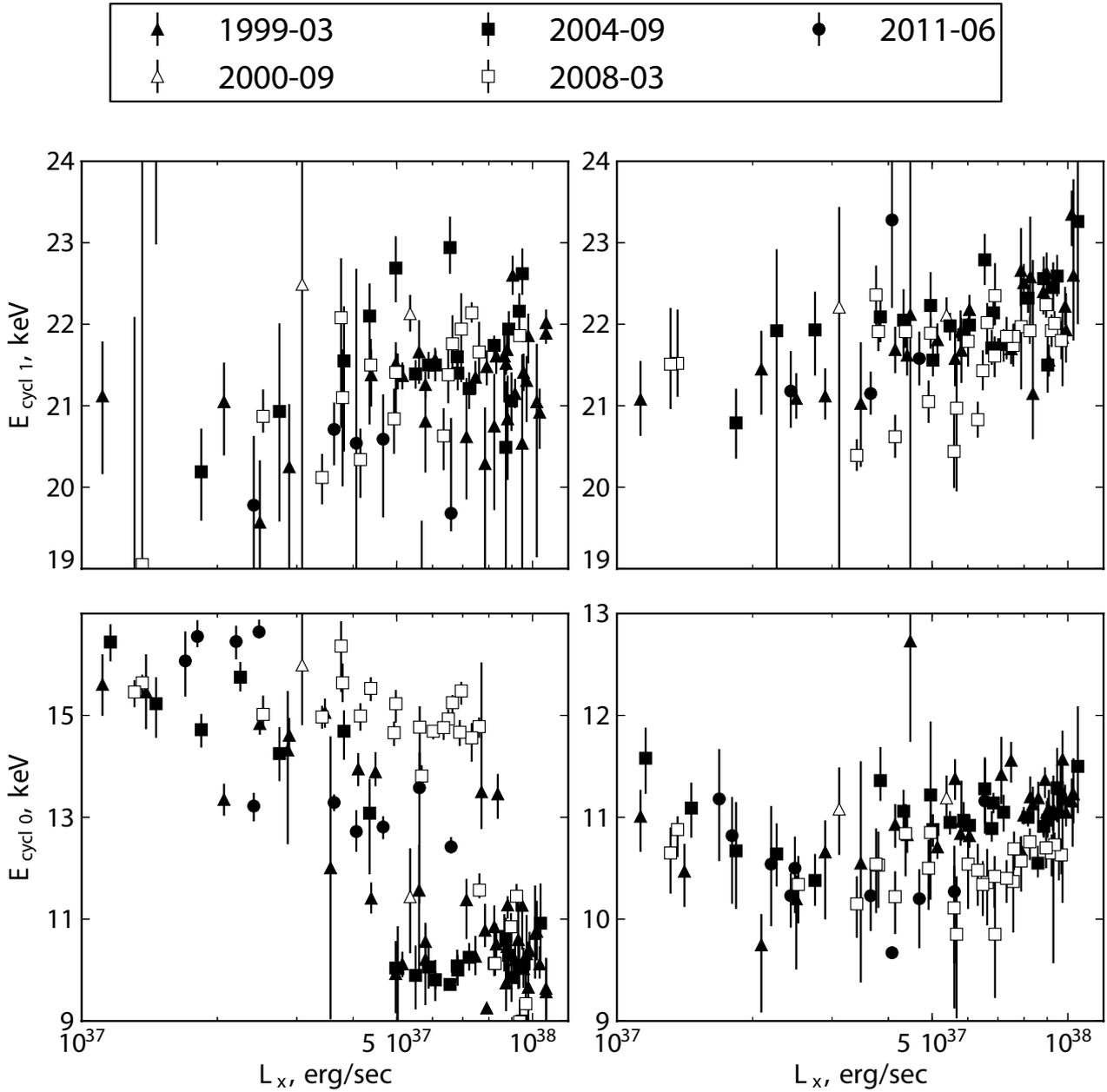}
    \caption{\label{pic:lum-cycles}
Energy of the fundamental and first harmonics of the cyclotron absorption line detected in the spectrum of the pulsar
4U 0115+63 versus its luminosity for different continuum models. The data were obtained from the results of PCA/RXTE
observations during bright outbursts (indicated by different symbols). The CSRF energies when the spectra were fitted
by a power law with a high-energy cutoff are presented in the left panels. An anticorrelation between the position of the fundamental
harmonic and luminosity (lower panel), which does not manifest itself for the first harmonic (upper panel), can be seen. The
energies of the cyclotron line harmonics when the spectra were fitted by a power law with a high-energy cutoff with the addition
of an emission line with a Gaussian profile at an energy of about 10 keV are presented in the right panels. The absence of an
anticorrelation between the cyclotron line energies and pulsar luminosity is obvious.
                }
\end{figure*}

Despite the enormous number of observations and
studies of cyclotron lines in the spectrum of the X-ray pulsar 4U 0115+63
(see above), several questions still remain open. One of the main questions
is whether the observed changes in the positions of
the fundamental cyclotron line with luminosity are
real \citep{nakajima2006,tsygankov2007,klochkov2011b} or
they are related to the chosen continuum model \citep{ferrigno2009,muller2013}.
Following the approach proposed by \cite{ferrigno2009} the authors
of the latter paper added an artificial
Gaussian component at energy $\simeq 10$ keV
to the phenomenological continuum model. In this
case, the anticorrelation between the cyclotron line
position and luminosity observed previously in the
RXTE and INTEGRAL data during the March/April
2008 outburst disappeared.

To test the conclusions reached by \cite{muller2013},
we investigated the spectra of the X-ray pulsar 4U 0115+63
and the dependence of the cyclotron
absorption line characteristics on the source's luminosity using
all of the available data obtained by the
RXTE/PCA spectrometer during intense outbursts
since 1995. Using the {\sc XSPEC} version v12.7.1b package,
we fitted the phase-averaged spectra in the 4-28 keV energy band with
two different models. The first model is a power law with a high-energy cutoff
multiplied by two cyclotron lines plus an iron line in
the form of a Gaussian. We chose this model as giving
the best $\chi^2$ values among all continuum models. The
energy and width of the fluorescent iron line were
fixed at the following values: $E = 6.4$ keV and $\sigma =
0.2$ keV; the iron line intensity was a free parameter.

When using this model, the anticorrelation between the cyclotron line energy
and luminosity that has already been found by
\cite{mihara2004,nakajima2006,tsygankov2007} manifests itself in all outbursts.
Obviously, the cyclotron
line energy lies within the range $11$-$12$ keV at high
luminosities and rises up to 16 keV with decreasing luminosity when the source's luminosity passes
through $L_x \sim (4-5)\times10^{37}$\ergs\ (see Fig. 3, the
lower left panel). The photon index $\Gamma$ for all fits lies
within the range $-0.1$-$0.5$ and shows no correlation with the line energy.
It is important to note that whereas the energy of the fundamental harmonic
changes significantly, the energy of the first harmonic
remains essentially constant (Fig. 3, the upper left panel).
Thus, it can be concluded that either we chose not quite a proper
continuum model or the first and fundamental harmonics are formed
in distinctly different regions.

No anticorrelation between the luminosity and position of
the fundamental cyclotron absorption line
harmonic is observed when using another continuum
model proposed by \cite{ferrigno2009}. This model
is the sum of a power law with a high-energy cutoff
and a broad emission feature with a Gaussian profile
energy $\approx 10$~keV and width $\approx 3$~keV.
As in the previous case, two cyclotron lines
and an iron line are added to this model.

The results of applying this model are presented in
Fig. 3 (right panels). It can be seen that when using
this model, the energy of the fundamental cyclotron
absorption line harmonic does not change practically with the
source's luminosity, and the relationship between
the energies of the fundamental and first harmonics
remains constant. This stems from the fact that the
motion of the emission line with a Gaussian profile
and the change in its intensity ”mask” the change in
the cyclotron absorption line energy. This assertion
was tested through the generation of spectra in the
{\sc XSPEC} package using the {\em fakeit} command and
the {\em (powerlaw * highecut + gauss) * cyclabs * cyclabs}
model followed by their fitting with the
{\em powerlaw * highecut * cyclabs * cyclabs} model. In this case, the
behavior of the energy of the fundamental cyclotron
absorption line harmonic in the generated spectra
closely followed its behavior when the observed spectra were fitted directly.

As has already been noted above, the Gaussian
emission line at energy $\sim10$ keV being added to the
continuum spectrum has no any physical meaning. Its inclusion in the model
formally improves the quality of the fit to individual spectra with confidence
levels $\approx2-3\sigma$ . Another factor that makes it difficult
to model the spectrum of 4U 0115+63 is that the
cutoff energy $E_\mathrm{cut}$ is very close to the energy of the
fundamental cyclotron absorption line harmonic and
it turns out to be very problematic to properly take into
account the influence of different factors.


Thus, the above results provide evidence for a
model strong dependence of the measured positions and evolution of the cyclotron
lines in the spectrum of the X-ray pulsar 4U 0115+63.
 A physically justified model
of the spectra for X-ray pulsars is needed to properly
describe its spectrum and to answer the question
about the evolution of the cyclotron absorption line
energy.

\subsection{Broadband Spectrum from INTEGRAL Data during the 2011 Outburst}
\begin{figure}
    \centering
    \includegraphics[width=1.0\columnwidth]{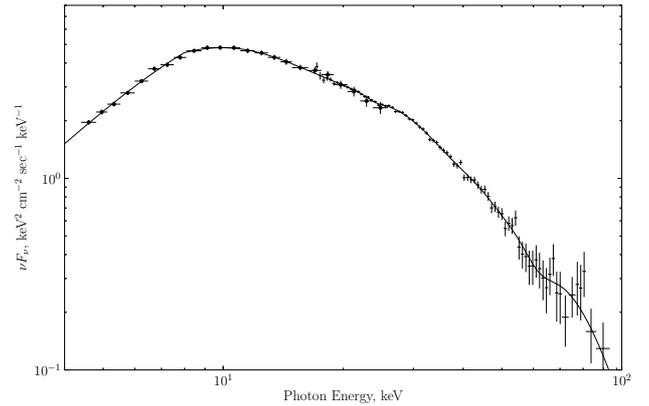}
    \caption{\label{pic:integral}
Fig. 4. Broadband energy spectrum of the X-ray pulsar 4U 0115+63 obtained from IBIS/ISGRI and JEM-X (INTEGRAL)
data in its bright state during the outburst in May-June 2011. The solid curve indicates the best-fit model; the circles and crosses
represent the JEM-X and IBIS/ISGRI measurements, respectively.
    }
\end{figure}

Figure 4 shows the spectrum of 4U 0115+63
in a wide energy range, 4-100 keV, obtained from
IBIS/ISGRI and JEM-X (INTEGRAL) data during
the 2011 outburst. During this outburst, the source
was observed by the INTEGRAL observatory several
times in states with different luminosities
(from $L_x\simeq7\times10^{37}$\ \ergs\ to $L_x\simeq10^{37}$\ \ergs).
For the presentation in Fig. 4, we used the data obtained in
revolutions 1061 and 1063 near the outburst
peak ($L_x\simeq7\times10^{37}$\ \ergs).

The data obtained were fitted by
the models, from a simple power law with a highenergy cutoff up
to the model with five harmonics of
the cyclotron absorption line. To align the ISGRI
and JEM-X spectra, we used a multiplicative factor
whose value reflects different calibrations of different
instruments but, nevertheless, turns out to be close to
unity. The spectral continuum parameters obtained
through our fitting for the bright state agree well
with those measured for the source being studied
by other observatories: photon index $\simeq0.2$,
$E_\mathrm{cut} \simeq 8.7$ keV,
and $E_\mathrm{fold} \simeq 11.6$ keV. The fundamental and
two first harmonics of the cyclotron absorption line
at energies $\simeq 11, 24$, and 35.6 keV, respectively, are
also detected in the source's spectrum at a high
confidence level ($>5\sigma$). Adding higher harmonics
also improves the quality of the fit, but the confidence
level of their detection turns out to be not so
high --- the confidence levels of the third and fourth
harmonics at energies 48.8 and 60.7 keV are $\sim 2\sigma$ and
$\sim 2.5 \sigma$, respectively. It is important to note that the
measured energy of the fourth harmonic agrees well
with the measurements made by \cite{muller2013}
based on RXTE data and differs significantly from
the energy measured for it from BeppoSAX data
\citep{ferrigno2009}. The component related to
the fluorescent iron line at energy 6.4 keV was also
included in the model. However, because of the wellknown problems
with the JEM-X response matrix at
energies close to its position, it is very difficult to make
any conclusions about its intensity and presence in
the source's spectrum.

In conclusion, note that adding a Gaussian line
at energy $\sim$10 keV to the model does not lead to
significant improvements in the quality of the fit to
the spectra obtained from INTEGRAL data. This
may be because the quality of the JEM-X data in
the 3-30 keV energy band is lower than that of the
PCA/RXTE data used in the previous section and
because the exposure time of each observations is
relatively short ($\sim$50 ks). Without this component,
the behavior of the energy of the fundamental cyclotron absorption
line harmonic almost exactly follows Fig. 3 (lower left panel) ---
at the source's high luminosity, the line is in the range $11.5$-$12$ keV;
as the luminosity decreases, it is shifted to $15.5$-$16$ keV.

\section{CONCLUSIONS}

A review of the available data on the X-ray pulsar
4U 0115+63, a member of the binary system with
the Be star V635 Cas, was carried out. The history
of X-ray outbursts from the source recorded over a
more than forty-year-long history of its observations
based on data from more than ten observatories and
instruments was presented. The evolution of the neutron
star spin frequency over the entire history of its
observations, including the results of recent
measurements made by the INTEGRAL observatory during
the 2004, 2008, and 2011 outbursts, was analyzed.

Based on the INTEGRAL data obtained at our
request during the 2011 outburst, we constructed
broadband spectra of the pulsar for various luminosity
levels. Five harmonics of the cyclotron absorption
line at energies $\simeq 11, 24, 35.6, 48.8$, and 60.7 keV were
shown to be detected in the source's spectrum, with
the energy of the fundamental harmonic changing
from $11$-$12$ keV in its bright state to $15.5$-$16$ keV
as its luminosity decreases.

To investigate the reality of the anticorrelation
between the fundamental harmonic energy and the
source's luminosity as well as its dependence on the
spectral continuum shape, we analyzed the spectra
based on all the available PCA/RXTE data obtained during bright outbursts
detected from the pulsar 4U 0115+63 in 1995 -- 2011. We showed that
including this additional component in the form of a
Gaussian broad emission line at energy $\simeq10$ keV
suggested previously by \cite{ferrigno2009} and
\cite{muller2013} in the model led to the disappearance
of the $E_{cyc,0}-L_x$ anticorrelation. Our modeling
showed that this effect results from the change in
emission line parameters that “masks” the possible
anticorrelation.
It is necessary to note that it is rather
problematic to physically justify the presence of an
additional emission component in the spectrum.
Another important factor that makes it difficult to model
the spectrum of 4U 0115+63 is that the cutoff energy
$E_\mathrm{cut} \simeq 8-9$ keV is very close both to the energy of
the fundamental cyclotron absorption line harmonic
$E_\mathrm{cyc,0} \simeq 11$ keV and to the energy of the artificially added
emission line at $\simeq 10$ keV. Collectively, all of these
factors do not allow the question about the presence
or absence of an $E_{cyc,0}-L_x$ anticorrelation for the
X-ray pulsar 4U 0115+63 to be unambiguously answered at present.
This requires a physically justified
model of the spectra for X-ray pulsars.

\section{ACKNOWLEDGMENTS}

We are grateful to E.M. Churazov who developed
the methods for analyzing data from the IBIS telescope of the INTEGRAL
observatory and for providing the software. We are also grateful to the referee
for helpful remarks and corrections. This work was
supported by the Russian Foundation for Basic Research
(project nos. 11-02-01328, 12-02-01265), the
''Origin, Structure, and Evolution of Objects in the
Universe'' Program of the Presidium of the Russian
Academy of Sciences, and the Program of the President of the Russian
Federation for Support of Leading Scientific Schools (project NSh-5603.2012.2).

\bibliography{4u0115}{}
\bibliographystyle{pazhbeng}

\end{document}